\documentstyle[aps,prl,epsfig]{revtex}
\begin{document}
\draft
\title{Observation of Pseudoscalar and Axial Vector Resonances in
 $\pi^- p \to K^+K^-\pi ^0 n$ at 18 GeV}
\author{G.S. Adams,$^1$ T. Adams,$^2$\footnote{Present address: Dept. of
Physics, Kansas State University, Manhattan, KS 66506.} Z. Bar-Yam,$^3$ 
J.M. Bishop,$^2$ V.A. Bodyagin,$^4$  
D.S. Brown,$^5$\footnote{Present address: Department of Physics, 
University of New Mexico, Albuquerque, NM 87131.} N.M. Cason,$^2$\\
S.U. Chung,$^6$ J.P. Cummings,$^1$ K. Danyo,$^6$ A.I. Demianov,$^4$ S. Denisov,$^7$ 
V. Dorofeev,$^7$ J.P. Dowd,$^3$\\ 
P. Eugenio,$^{3}$\footnote{Present address: Dept. of Physics, 
Carnegie Mellon University, Pittsburgh, PA 15213.}
X.L. Fan,$^5$ A.M. Gribushin,$^4$  R.W. Hackenburg,$^6$ 
M. Hayek,$^3$\footnote{Permanent address: Rafael, Haifa, Israel.} J. Hu,$^1$  
E.I. Ivanov,$^2$ D. Joffe,$^5$\\
I. Kachaev,$^7$ W. Kern,$^3$ E. King,$^3$ 
O.L. Kodolova,$^4$ V.L. Korotkikh,$^4$ M.A. Kostin,$^4$ 
J. Kuhn,$^1$ V. Lipaev,$^7$\\
J.M. Losecco,$^2$ M. Lu,$^1$ J.J. Manak,$^2$ 
J. Napolitano,$^1$ M. Nozar,$^1$\footnote{Present address: Thomas
Jefferson National Accelerator Facility, Newport News, VA 23606.} 
C. Olchanski,$^6$ A.I. Ostrovidov,$^{4}$\\
T.K. Pedlar,$^5$\footnote{Present address: Laboratory for Nuclear Science,
Cornell University, Ithaca, NY 14853.} A. Popov,
$^7$ D. Ryabchikov,$^7$ A.H. Sanjari,$^2$
L.I. Sarycheva,$^4$ K.K. Seth,$^5$ N. Shenhav,$^{3\ddag}$\\
X. Shen,$^{5,8}\footnote{Permanent address: Institute of High Energy
Physics, Bejing, China 100039.}$
W.D. Shephard,$^2$ N.B. Sinev,$^4$ D.L. Stienike,$^2$
S.A. Taegar,$^2$\footnote{Present address: Dept. of Physics,
University of Arizona, Tucson, AZ 85721.} D.R. Thompson,$^2$
A. Tomaradze,$^5$\\
I.N. Vardanyan,$^4$ D.P. Weygand,$^{8}$ D. White,$^1$
H.J. Willutzki,$^6$ A.A. Yershov,$^4$\\[0.3cm]
}
\address{
$^1$Department of Physics, Rensselaer Polytechnic Institute, Troy, New York 12180\\
$^2$Department of Physics, University of Notre Dame, Notre Dame, Indiana 46556\\
$^3$Department of Physics, University of Massachusetts Dartmouth, North 
Dartmouth, Massachusetts 02747\\
$^4$Nuclear Physics Institute, Moscow State University, Moscow, Russia 119899\\
$^5$Department of Physics, Northwestern University, Evanston, Illinois 60208\\
$^6$Brookhaven National Laboratory, Upton, New York 11973\\
$^7$Institute for High Energy Physics, Protvino, Russia 142284\\
$^8$Physics Dept., Thomas Jefferson National Accelerator Facility,
Newport News, VA 23606\\
}
\maketitle
\vspace{0.5cm}\centerline{(E852 Collaboration)}
\date{\today}
\begin{abstract}
The number of pseudoscalar mesons in the mass range from 1400 to 1500 MeV$/c^2$ has
been a subject of considerable interest for many years, with several
experiments having presented evidence for two closely spaced states. A 
new measurement of the reaction $\pi^- p \to K^+K^-\pi^0 n$ has been 
made at a beam energy of 18 GeV. A partial wave analysis of the 
$K^+K^-\pi^0$ system shows evidence for three pseudoscalar resonances, 
$\eta(1295)$, $\eta(1416)$, and $\eta(1485)$, as well as two 
axial vectors, $f_1(1285)$, and $f_1(1420)$. Their observed masses, widths 
and decay properties are reported.  No signal was observed for $C(1480)$, 
an $I^G J^{PC} = 1^+ 1^{--}$ state previously reported in $\phi\pi^0$ decay.
\end{abstract}


\twocolumn

\par
{\hskip 0cm}
A long history of confusion surrounds the pseudoscalar meson spectrum \cite{pdg,jim}.
Results from previous experiments
\cite{rath,birman,manak,lee,oblx1,oblx2,oblx3,jsi1,jsi2,jsi3,ext1,ext2,ext3,ext4,ext5,ext6}
carried out over the 
past four decades, including peripheral production with pion beams, 
$p\bar p$ annihilation, and $J/\psi$ radiative decay, indicate that
there may be three isosinglet pseudoscalar states in the narrow mass range from 1250 to 
1500 MeV$/c^2$.
However only two states are expected in the standard $J^{P} = 0^-$ nonet.

\par
{\hskip 0cm}
The lowest pseudoscalar, $\eta(1295)$, was first observed in a PWA 
analysis of the $\eta \pi\pi$ system \cite{e1295-1} and later confirmed 
by other experiments \cite{e1295-2,e1295-3,e1295-4,e1295-5}. However
the $\eta(1295)$ was not observed in some reactions where the production of 
pseudoscalars is allowed, such as $\gamma\gamma$ collisions \cite{aihara}, 
central production \cite{armstrong,barberis} and $J/\psi$ decays
\cite{jsi1,jsi2,jsi3}. 
Therefore further confirmation 
of this state is desirable. 

\par
{\hskip 0cm}
The central issue in the higher mass region is sometimes referred to 
as the $E/\iota$ puzzle. There was a longstanding question whether the iota, seen in
radiative $J/\psi$ decay \cite{ext2,ext3}, and the E meson (now called $f_1(1420)$)
were the same state or two distinct resonances. Peripheral production experiments at 
Brookhaven National Laboratory showed a complex spectroscopy in this mass 
region \cite{rath,birman,mpsconf,kref}.  

\par
{\hskip 0cm}
One of these, experiment E771, used pion, kaon 
and antiproton beams to study the production of different states \cite{birman,mpsconf,kref}.
The kaon data showed that the $f_1(1285)$ has very little strange-quark content \cite{kref}.  
Evidence was also obtained for a $J^{PC} = 1^{+-}$ resonance at about 1380 MeV$/c^2$ \cite{kref},
below the $K^* \overline K$
threshold.  
Those
results were consistent with data obtained by the LASS collaboration \cite{lass}. They 
also observed $J^{PC} = 1^{++}$ states at 1420 and 1500 MeV$/c^2$ \cite{birman}, and
two pseudoscalar states \cite{mpsconf,kref} in the 1400 MeV$/c^2$ mass region,
one decaying mainly to 
$K^* \overline K$ and the other decaying primarily to $a_0(980)\pi$.  
Evidence was also presented for an additional pseudoscalar at 1515 MeV$/c^2$ 
appearing only in $a_0(980)\pi$ \cite{mpsconf}.

\par
{\hskip 0cm}
In their 
latest publication \cite{jsi2} the Mark III group conclude that their data require 
two $J^{P} = 0^-$ resonances: one at 1416 MeV$/c^2$ decaying to $a_0(980)\pi$ 
and the other at 1490 MeV$/c^2$ decaying to $K^* \overline K$. 
The DM2 collaboration \cite{jsi3} present quite a different picture: a $0^-$ resonance 
at 1421 MeV$/c^2$ decaying to $K^* \overline K$ and a $0^-$ resonance at 1459 
decaying to $a_0(980)\pi$. The Obelix group recently reported results 
for $p \bar p \to (K^{\pm}K_S\pi^{\mp})\pi^+\pi^-$ at rest \cite{oblx1,oblx2,oblx3}. 
They observe two $0^-$ resonances in the $(K\overline K \pi)$ final state, 
one at 1405 MeV$/c^2$ decaying directly to $(K\overline K \pi)$ and a 
broader one at 1500 MeV$/c^2$ decaying primarily to $K^* \overline K$. The PDG \cite{pdg}
concludes
that most 
of the evidence now supports two $0^-$ states in the range 1.4-1.5 GeV$/c^2$, 
but their decay properties are poorly determined.  The present 
experiment was designed to clarify this situation by obtaining a large sample 
of data for the reaction $\pi^- p \to K^+K^-\pi^0 n$.  The task is a challenging one because 
of the large number of overlapping resonances in this mass region, and 
the presence of the $K^* \overline K$ threshold 
at about 1.4 GeV/$c^2$.

\par
{\hskip 0cm}
One controversial state is the $C(1480)$, which was found by the Lepton--F 
group in $\phi(1020)\pi^0$ decay \cite{c1480-1}. The mass and width of this 
resonance were measured to be $M = 1480\pm40$ MeV$/c^2$, and 
$\Gamma = 130\pm60$ MeV$/c^2$, respectively, with $I^G J^{PC} = 1^+ 1^{--}$. 
Because it has a large production cross section in $\pi^- p \to (K^+K^-\pi^0) n$ 
and its decay into $\omega \pi^0$ was not observed, it has been suggested that 
this state is not a conventional meson \cite{c1480-2}.  The PDG now lists $C(1480)$
under $\rho(1450)$, a state previously associated with non-strange decay modes\cite{pdg}.

\par
{\hskip 0cm}
The present experiment (E852) was carried out during the 1997 running 
period of the Alternating Gradient Synchrotron at Brookhaven National 
Laboratory. A diagram of the experimental apparatus is shown in Fig. 1. 
A Cherenkov-tagged $\pi^-$ beam of momentum 18.3 GeV$/c$ was incident on a 60 cm 
liquid hydrogen target within the Multi-Particle Spectrometer. The 
spectrometer central field was set to 0.75 Tesla. The target was 
surrounded by a four-layer cylindrical wire chamber (TCYL)
which was used to reject events that included a large-angle  
charged particle \cite{mps1}. A segmented $CsI$ detector was used to
reject events with large-angle photons \cite{mps2}.  

\par
{\hskip 0cm}
The downstream part of the magnet was equipped with 
six multi-plane drift-chamber modules (DC1-6) \cite{mps3} for charged-particle 
tracking. Three proportional wire chambers (TPX1-3) allowed the acquisition 
system to trigger on events having two charged tracks. A 3045-element 
lead-glass calorimeter (LGD) \cite{mps4} was used to detect forward 
photons, and a 96-segment threshold Cherenkov counter was used to distinguish 
charged kaons from pions. The Cherenkov radiator was Freon 
114 at atmospheric pressure, giving a diffractive index of 1.00153.
The corresponding pion (kaon) threshold is at 2.52 (8.93) GeV$/c$.

\par
{\hskip 0cm}
The trigger for the present data required two forward-going charged tracks, energy
deposition in the LGD, and no charged recoil track. A total of 18.5 million events
of this type were recorded. After calibration and track fitting the events were
filtered by requiring:
\begin{itemize}
\item[1.]a fully reconstructed beam track; 
\item[2.]two forward tracks, one positive and one negative, both 
identified as kaons by the Cherenkov detector;
\item[3.]two energetic clusters in the active region of the LGD, both 
identified as neutral particles; 
\item[4.]a reaction vertex within the target volume;
\item[5.]energy deposition $< 40$ MeV in the $CsI$ detector.
\end{itemize} 

\par
{\hskip 0cm}
In the final step of the data selection a kinematic fit was made to 
the $\pi^- p \to K^+ K^- \pi^0 n$ reaction hypothesis using the full 
covariance matrix from track and vertex reconstruction\cite{squaw}. Events with 
confidence level less than six percent were rejected.  This was done to 
remove events with low-energy photons that were not detected by the 
apparatus. Approximately 34,000 events survived the above selection 
criteria. Fig. 2 shows the invariant mass distributions for these 
exclusive events. The most striking features of the data are two peaks in the 
$K^+K^-\pi^0$ mass spectrum at about 1.3 and 1.45 GeV$/c^2$.
 The two-particle mass spectra show clear signatures 
for $\phi(1020)$ and $K^*(892)$ excitation.

\par
{\hskip 0cm}
A partial wave analysis (PWA) of the present data was made in the isobar 
model \cite{pwa,bnlint}. 
The mass of the three-meson final state was binned in 20 MeV$/c^2$ intervals  
and independent fits were performed on the data in each bin.
The final state was represented as a 
sequence of interfering two-body intermediate states. An initial decay
of a parent meson into a $K\overline K$ or $K\pi^0$ intermediate resonance 
(isobar) and an unpaired pion or kaon, followed by the subsequent decay of 
the isobar, populates the $K^+ K^- \pi^0$ final state. Each partial wave is 
characterized by the quantum numbers $J^{PC}[isobar]LM^{\epsilon}$, where 
$J^{PC}$ are the spin, parity and $C$-parity of the partial wave, $M$ is the 
absolute value of the spin projection on the beam axis, $\epsilon$ is the 
reflectivity (corresponding to the naturality of the exchanged particle), 
and $L$ is the orbital angular momentum between the isobar and the unpaired 
particle.  Both nucleon spin-flip and non-flip amplitudes were included.

\par
{\hskip 0cm}
The experimental acceptance was determined by means of a Monte Carlo 
simulation, which was then incorporated into the PWA normalization for 
each partial wave. In the description of the isobars a coupled-channel 
Flatt\'e parametrization \cite{a01} was used for the
$a_0(980)$.  The parameters of this function were set to the values
used in previous partial wave analyses \cite{birman,lee}.  
A scattering length function was used to
describe $(K\pi)_S$, the $(K\pi)$ S-wave interaction \cite{kpi-s}.
The other isobars, $K^*(892)$, and $\phi(1020)$, were  modeled by
relativistic Breit-Wigner shapes with masses extracted from the
PDG \cite{pdg}.  Experimental resolution made a significant
contribution to the measured width of the $\phi(1020)$ so in that case 
the observed width (9.5 MeV$/c^2$) was used in the PWA analysis.

\par
{\hskip 0cm}
In the partial wave analysis numerous fits with different wave sets were 
performed to determine the minimum set of waves that gave a good description 
of the angular distributions and the two-particle mass distributions.  All 
waves were included as $C$-parity eigenstates \cite{bnlint} (eg. $K^{*+}K^- \pm K^{*-}K^+$).
Partial waves with $J<4$ were included for $K^*$, $a_0$ and $\phi$ isobars
as well as $(K\pi)_S$.
 The extended maximum likelihood method was used to determine
the goodness-of-fit. 

\par
{\hskip 0cm}
The largest 
contributions came from pseudoscalar and axial vector waves, with natural-parity 
exchange dominating the results. Table I lists all partial waves used in the 
final fit. Waves with $J^{PC} = 2^{++}$ and $1^{--}$ were employed to account 
for the presence of strong tails from the $\rho(1700)$ and other high-lying 
states. Both of these were needed in the fits.
The $1^{+-} K^*\overline K$ S-wave was included to allow for possible 
excitation of the $h_1(1380)$. A non-interfering isotropic background wave was 
included at each mass bin to simulate the cumulative effect of numerous small 
waves that were omitted from the fit.  

\par
{\hskip 0cm}
The final fit was performed on 20,000 events having momentum transfer
 $|t| > 0.1$ GeV$^2$.  This cut reduced 
the contributions from $\phi\pi$ decay (see Fig. 2), 
and also the strength of the $1^{--}$ waves.

\par
{\hskip 0cm}
No $(K\pi)_S \overline K$ waves were included in the final fit. Those waves 
were found to be ambiguous with the $a_0\pi$ S-waves and with the 
background wave. Fits which included 
both $(K\pi)_S$ and $a_0$ waves exhibited large interference that varied 
wildly in strength at adjacent $M(KK\pi)$ mass bins. Therefore one cannot 
exclude the possibility of some undetected $(K\pi)_S \overline K$ strength in 
the present analysis.

\par
Different wave sets were used to describe the data above and below the
$K^* \overline K$ threshold.
Fits below 1.375 GeV$/c^2$ did not include $K^* \overline K$ waves.   
The main results of the mass-independent PWA fit are shown in Figs. 3 and 4. 
The low-mass spectrum shows large contributions from $J^{PC}$ = $1^{++}$ and 
$0^{-+}$ waves. 

\par
Mass-dependent fits to the intensities and phase 
differences were made by least-squares
minimization for three separate wave pairs.
 The phase information is particularly important for
determining the resonant content of the mass-independent results.
These mass-dependent fits used linear combinations of relativistic Breit-Wigner 
poles with mass-dependent widths and Blatt-Weisskopf barrier factors \cite{etapi}. 
The details of this procedure are given below.

\par
The first fit was to the $a_0\pi^0$ waves below 1.36 GeV$/c^2$. Those results, depicted 
in Fig. 3 (a-c), show clear excitation of $\eta(1295)$ and $f_1(1285)$. The 
resonance parameters from this fit are tabulated in Table II. Contributions from 
experimental resolution (about 10 MeV$/c^2$) were removed from 
the listed widths.  The errors given in the table include statistical contributions 
from the mass-dependent fit (first entry), and systematic errors from the PWA fits
(second entry). The latter contribution was estimated by comparing the
results of several mass-dependent fits which used different PWA results.
In those fits the mass binning and wave sets were varied. 

\par
With the exception of the $f_1(1285)$ width, the measured parameters of the 
$\eta(1295)$ and $f_1(1285)$ are in good agreement with the average values 
reported by the PDG \cite{pdg}.
For the $f_1(1285)$ we observe a width $\Gamma = 45 \pm 9 \pm 7$ MeV$/c^2$, 
which is consistent with the latest measurement from GAMS \cite{e1295-4} but 
larger than the PDG value of $24.0 \pm 1.2$ MeV$/c^2$. 
Fitting only the intensity function in this case (data in Fig. 3b)
yields a much smaller width, $\Gamma = 23\pm5$ MeV$/c^2$. Thus 
it is the interference with $\eta(1295)$ that is important in our
determination of the width.

\par
{\hskip 0cm}
Above 1.38 GeV$/c^2$ several overlapping states show interesting structure, as indicated 
in Fig. 4. 
Pseudoscalar waves show up prominently in $a_0\pi^0$ and $K^* \overline K$, 
while the most important axial vector wave is in $K^* \overline K$ decay. The 
pseudoscalar mass spectrum for $a_0\pi^0$ decay is markedly different from 
that for $K^* \overline K$. The former shows a narrow peak in the intensity 
distribution while the latter shows strength over a much wider mass range.

\par
{\hskip 0cm}
The results of the mass-independent PWA fits for this mass region were interpreted 
in two steps. In the first step the intensities and relative phase of the 
most prominent pseudoscalar $a_0\pi^0$ and axial vector $K^* \overline K$ 
waves were fitted with two Breit-Wigner poles.  The results of that fit are 
given in Fig. 4 (a-c) and in Table II.  They are consistent with previous 
observations of $f_1(1420)$ and a low-mass component of $\eta(1440)$\cite{pdg}, seen 
here at a mass of 1416$\pm4\pm2$ MeV$/c^2$.  In the present work this state will 
be labeled $\eta(1416)$.  The mass and width of the 
$f_1(1420)$ agree with the average values 
reported by the PDG \cite{pdg}.

\par
{\hskip 0cm}
In the second step of the mass-dependent analysis the intensities and relative 
phase of the prominent pseudoscalar and axial vector $K^* \overline K$ waves 
were fitted to three resonance poles. The parameters of the $f_1(1420)$ and  
$\eta(1416)$ were held constant at the values determined above (see Table II), 
while the mass and width of a third resonance were varied to minimize $\chi^2$.  
The results of this fit are given in Fig. 4 (d-f) and in Table II. They show a mass 
and width for the high-mass component of $\eta(1440)$ equal to 1485$\pm8\pm5$ MeV$/c^2$ 
and 98$\pm18\pm3$ MeV$/c^2$, respectively. This state will be labeled $\eta(1485)$.  
For this fit $\chi^2/d.o.f$ was 1.07 (for 16 degrees of freedom).  
Fits were also made with the parameters of the 
lower states unconstrained.  Those results were in good agreement with the values 
given in Table II, but with larger error bars.  
Previous PWA analyses have determined that the $K^+  \overline K^0 \pi^-$ 
spectrum is dominated by positive $G$-parity excitation \cite{birman,e1295-5},
so one is justified in assigning isospin $I=0$ to the positive $C$-parity
states discussed above.  

\par
The present results agree with those from earlier $\pi^- p$ \cite{rath}, 
$p \bar p$ \cite{oblx1}, and $J/\psi$ radiative decay \cite{jsi2} experiments; 
two pseudoscalar resonances were observed, with the lower state appearing 
clearly in $a_0\pi^0$ decay and the upper one strongly populating only the 
$K^* \overline K$ spectrum.
In the present measurements $\eta(1416)$ was 
observed in both $a_0\pi^0$ and $K^* \overline K$ decay.  
In order to test the significance of this result a new fit of the intensities and 
relative phase of the waves depicted in Fig. 4 was made. In that fit a single 
resonance was assumed in the pseudoscalar $K^* \overline K$ wave. This resulted 
in the width of the resonance increasing to about 300 MeV$/c^2$, and $\chi^2/d.o.f$ 
increasing from 1.07 to 1.38 (for 17 degrees of freedom), which demonstrates that
 the two-pole fit offers an 
improved description of the data.

\par
{\hskip 0cm}
No strong signals were observed for 
excitation of the $f_1(1510)$.  As one can see in Fig. 4, the $J^{PC}$ = $1^{++}$ 
$K^* \overline K$ intensity shows a small bump at 1.5 GeV$/c^2$ but no clear phase 
motion with respect to the $\eta(1485)$.

\par
\hspace{0cm}
The decay properties of the $\eta(1416)$ are of particular interest for 
determining the structure of the state.  
From our measurements we obtain the ratio of branching fractions,
$$ \frac {BR[\eta(1416) \to K^* \overline K, K^* \to K\pi^0]} 
{BR[\eta(1416) \to a_0\pi^0,~a_0 \to K^+ K^-]}= 0.4 \pm 0.1$$
When all charge states are included this leads to the ratio of total branching fractions, 
$$ \frac {BR[\eta(1416) \to K^* \overline K+c.c.]} 
{BR[\eta(1416) \to a_0\pi]}= 0.084\pm0.024$$
The quoted error is statistical in nature.  This ratio is reduced by about a factor of 
two when pseudoscalar 
$(K\pi)_S \overline K$ waves are included in the fits.  
The above values are consistent with early results obtained by Obelix 
when similar wave sets were used \cite{oblx3}, but is much smaller than those obtained in later
Obelix experiments \cite{oblx1,oblx2}.

\par
{\hskip 0cm}
The $\eta(1416)$ and $\eta(1485)$ states were both observed to decay 
to $K^* \overline K$. The ratio of production rate times branching fraction
 for these two states is,   
$$ \frac {R[\eta(1416) \to K^* \overline K]} 
{R[\eta(1485) \to K^* \overline K]}= 0.16\pm0.04$$
Previous $\pi^- p$ experiments were unable to detect the 
$\eta(1416)$ contribution to the $K^* \overline K$ spectrum \cite{rath,birman,mpsconf}.

\par
{\hskip 0cm}
In principle one should be able to use the measured widths and decay rates of 
$\eta(1416)$ and $\eta(1485)$ to determine their valence structure.  Unfortunately the 
present state of hadron theory does not allow this. In the $^3P_0$ model 
one expects a pseudoscalar $s \overline s$ meson to have a narrow width 
($< 100$ MeV$/c^2$) and a dominant $K^* \overline K$ decay branch \cite{barnes}.  
However both $\eta(1416)$ and $\eta(1485)$ have widths that fall within 
acceptable limits, and both decay to $K^* \overline K$. 
The fact that $\eta(1485)$ decays primarily by kaon emission suggests that it is the
better $s \overline s$ candidate, but further theoretical work is needed 
before an identification can be made. 

\par
{\hskip 0cm}
Another important spectroscopic issue is the 
existence of $C(1480)$. That state was first identified in a production 
experiment similar to the present one \cite{c1480-1}. It was observed in
$\phi(1020)\pi^0$ decay with a mass and width $M = 1480\pm40$ MeV$/c^2$, and 
$\Gamma = 130\pm60$ MeV$/c^2$.

Subsequently it was not seen in several other experiments, including $pp$ central
production \cite{armstrong} and $p\bar p$ annihilation at rest \cite{reif}.
A \(\phi(1020)\) side-band study and partial wave analysis were used
to search for such a state in the present data.  Both studies were carried 
out using the full data set without a $t$ cut.

\par
{\hskip 0cm}
In the $K^+K^-$ effective mass spectrum a clear peak is seen at the mass 
of the $\phi$ meson (Fig. 2b). The intensity peaks at $1020.3\pm0.4$ 
MeV$/c^2$ and has a width of $9.5\pm1.0$ MeV$/c^2$. The width is mainly 
from the instrumental resolution of the detector. For the side-band study the 
two-kaon mass spectrum was divided into three bins, each 10 MeV$/c^2$ wide, and 
$K^+K^-\pi^0$ mass distributions were produced for each bin. Those spectra 
are shown in Fig. 5. Fig. 5b shows the $K^+K^-\pi^0$ effective mass spectrum 
for those events in the $\phi$ signal region. Background was removed from this 
spectrum by subtracting the average number of events in the adjacent two-kaon 
mass intervals. The resulting background-subtracted mass spectrum is shown in 
Fig. 5d. No peaks can be discerned in the 
spectrum.

\par
{\hskip 0cm}
A more definitive test was made by performing additional PWA fits with 
$\phi \pi^0$ $S$ and $P$ waves added to the wave set given in Table I. No peaks 
were observed in the resulting mass spectra.  Since the $C(1480)$ was believed 
to be a vector meson, these fits included all allowed $M^{\epsilon}$ values 
for $1^{--} \phi \pi^0$. Fig. 6 shows the contribution from the wave with the 
largest intensity ($M^{\epsilon} = 0^-$).

\par
{\hskip 0cm}
The present results indicate no strong excitation of any $\phi \pi^0$ waves 
and negligible resonance strength for $1^{--} \phi \pi^0$ waves in particular.  
The previous identification of 
$C(1480)$ cannot be confirmed. 

\par
{\hskip 0cm}
In summary, a partial wave analysis of the mesons produced in the reaction 
$\pi^- p \to K^+K^-\pi^0 n$ was performed.  The intensity and phase of the 
resulting waves show clear excitation of several previously identified states: 
$f_1(1285)$, $\eta(1295)$, $\eta(1416)$, $f_1(1420)$, and $\eta(1485)$. The 
existence of three low-mass pseudoscalars is confirmed. The $\eta(1416)$ and 
$\eta(1485)$ are distinguished by their decay properties; $\eta(1416)$ decays 
primarily to $a_0\pi^0$ but has a small $K^* \overline K$ branch as well.  The 
$\eta(1485)$ was observed only in $K^* \overline K$ decay. 
$C(1480)$, a state previously identified 
in $K\overline K\pi$ decay, was not observed. 

\par
{\hskip 0.3cm}
We are grateful to the members of the MPS group for their outstanding efforts 
in running this experiment. This research was supported in part by the U.S. 
Department of Energy, the National Science Foundation, and the Russian State 
Committee for Science and Technology.

\begin{center}

\begin{figure}[htbp]
\center{\mbox{\epsfig{file=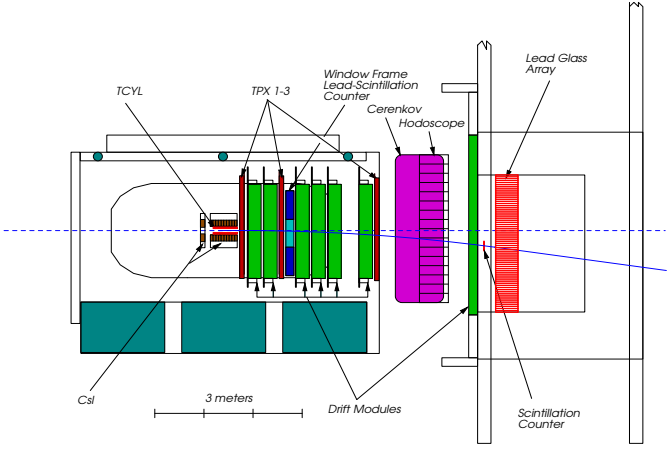,height=2.5in,width=3.15in}}}
\vspace{0.5cm}
\caption{Schematic diagram of the experimental apparatus.}
\end{figure}

\begin{figure}[htbp]
\center{\mbox{\epsfig{file=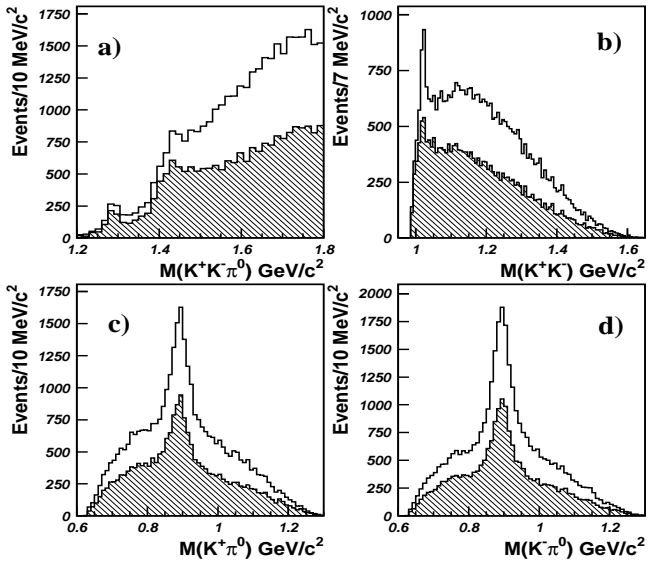,height=2.9in,width=3.35in}}}
\vspace{0.5cm}
\caption{Mass spectra for (a) $K^+K^-\pi ^0$, (b) $K^+K^-$,
(c) $K^+\pi^0$, and (d) $K^-\pi^0$. The open areas are all events and 
the shaded areas are events 
with $|t| > 0.1$ GeV$^2$.}
\end{figure}

\begin{figure}[htbp]
\center{\mbox{\epsfig{file=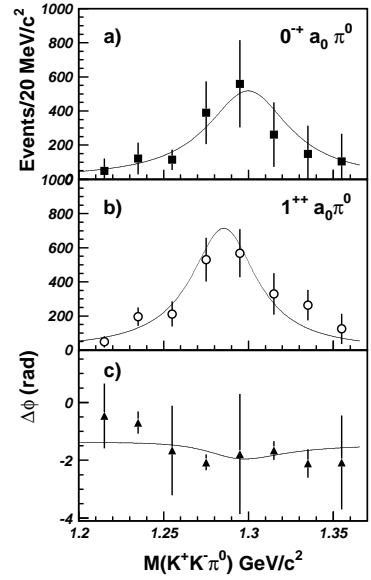,height=3.0in}}}
\vspace{0.5cm}
\caption{Results of mass-independent fits (points) and 
mass-dependent fits (lines) for 
(a) $0^{-+}$ $[a_0(980)]$ $S0^+$ intensity,
(b) $1^{++}$ $[a_0(980)]$ $P0^+$ intensity,
(c) phase difference between (a) and (b).}
\end{figure}

\begin{figure}[htbp] 
\center{\mbox{\epsfig{file=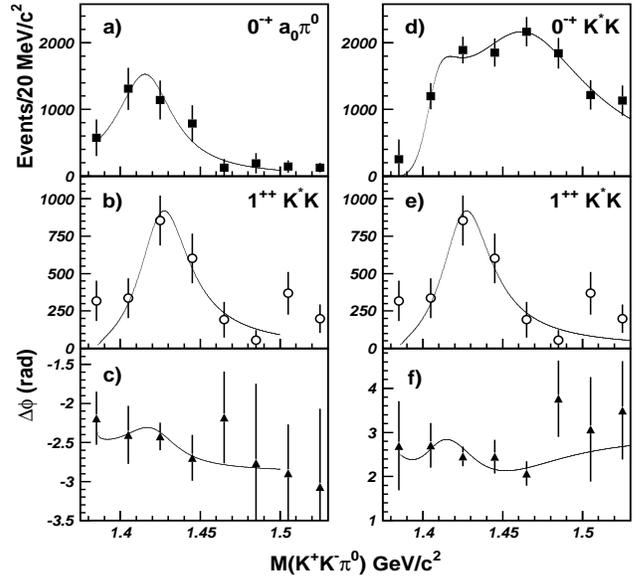,height=3.0in,width=3.25in}}}
\vspace{0.5cm}
\caption{Results of mass-independent fits (points) and 
mass-dependent fits (lines) for 
(a) $0^{-+}$ $[a_0(980)]$ $S0^+$ intensity,
(b) $1^{++}$ $[K^*(892)]$ $S0^+$ intensity, and  
(c) phase difference between (a) and (b),
(d) $0^{-+}$ $[K^*(892)]$ $P0^+$ intensity,
(e) $1^{++}$ $[K^*(892)]$ $S0^+$ intensity, and
(f) phase difference between (d) and (e).} 
\end{figure}

\begin{figure}[htbp]
\center{\mbox{\epsfig{file=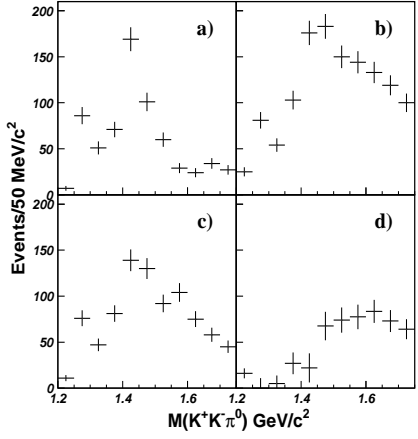,height=2.25in,width=2.15in}}}
\vspace{0.5cm}
\caption{$\phi(1020)$ side-band study. The $K^+K^-\pi^0$ effective mass 
for events with
(a) $1.0 < M(K^+K^-) < 1.01$ GeV$/c^2$, 
(b) $1.015 < M(K^+K^-) < 1.025$ GeV$/c^2$, 
(c) $1.03 < M(K^+K^-) < 1.04$ GeV$/c^2$, and
(d) the background-subtracted signal.}
\end{figure}

\begin{figure}[htbp]
\center{\mbox{\epsfig{file=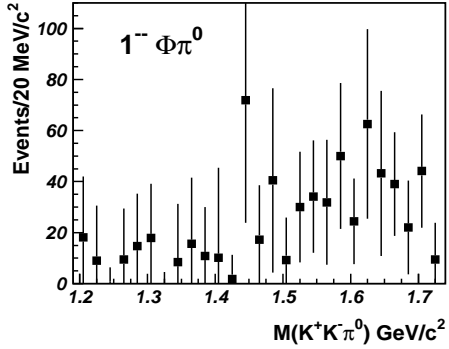,height=1.8in,width=2.3in}}}
\vspace{0.5cm}
\caption{Intensity of the $1^{--} [\phi(1020)]P0^-$ wave.}
\end{figure}

\begin{table}[htpb]
\label{table1}
\caption{Partial waves used in the amplitude analysis.  Note that the $K^* \overline K$ partial waves
were used only for masses greater than 1.375 GeV$/c^2$}
\vspace{0.35cm}
\begin{tabular}{|cccc|}
$J^{PC}$ & M$^\epsilon$ & L & Decay Mode\\
\hline\hline
$0^{-+}$ & $0^+$ & S & $a_0(980) \pi^0$\\
         &       & P & $K^*(892) \overline K$\\
\hline
$1^{++}$ & $0^+$ & S & $K^*(892) \overline K$\\
         &$0^+$, $1^{\pm}$  & P & $a_0(980) \pi^0$\\
\hline
$1^{+-}$ & $0^+$ & S & $K^*(892) \overline K$\\
\hline
$1^{--}$ & $0^-$ & P & $K^*(892) \overline K$\\
\hline
$2^{++}$ & $0^-$, $1^+$ & D & $K^*(892) \overline K$\\
\end{tabular}
\end{table}

\begin{table}
\label{table2}
\caption{Resonance parameters and decay modes of the observed states.
Statistical errors are listed first, followed by systematic errors.}
\vspace{0.35cm}
\begin{tabular}{|cccc|} 
Resonance & M (MeV$/c^2$)& $\Gamma$ (MeV$/c^2$) & Decay Modes \\
\hline\hline
$f_1(1285)$ & $1288\pm4\pm5$& $45\pm9\pm7$ & $a_0\pi^0$ \\
\hline
$\eta(1295)$ & $1302\pm9\pm8$& $57\pm23\pm21$ & $a_0\pi^0$ \\
\hline
$\eta(1416)$ & $1416\pm4\pm2$& $42\pm10\pm9$ & $a_0\pi^0,K^*\overline K$\\
\hline
$f_1(1420)$ & $1428\pm4\pm2$& $38\pm9\pm6$ & $K^*\overline K$\\
\hline
$\eta(1485)$ & $1485\pm8\pm5$& $98\pm18\pm3$ & $K^*\overline K$\\
\end{tabular}
\end{table}

\end{center}

\end{document}